\documentclass[showpacs,amsmath,amssymb,twocolumn,superscriptaddress,floatfix,prl]{revtex4-2}
\usepackage[dvips]{graphicx}
\usepackage{enumerate}
\usepackage{braket}
\usepackage{epsfig}
\usepackage{xcolor}
\usepackage[T1]{fontenc}
\usepackage{fullpage}
\usepackage{amsthm,amsfonts,amscd,mathrsfs,xspace,framed}
\usepackage{color}
\usepackage{setspace}
\usepackage{url}
\usepackage{enumitem}
\usepackage{soul}
\usepackage{subfiles}
\usepackage[subpreambles=true]{standalone}
\usepackage{import}

\makeatletter
\def\maketitle{
\@author@finish
\title@column\titleblock@produce
\suppressfloats[t]}
\makeatother

\begin{document}

\title{Superadditive Communication with the Green Machine as a Practical Demonstration of Nonlocality without Entanglement}

\author{Chaohan Cui}
\affiliation{Department of Electrical and Computer Engineering, The University of Maryland, College Park, Maryland 20742, United States}
\affiliation{James C. Wyant College of Optical Sciences, The University of Arizona, Tucson, Arizona 85721, United States}
\author{Jack Postlewaite}
\affiliation{Department of Electrical and Computer Engineering, The University of Maryland, College Park, Maryland 20742, United States}
\author{Babak N. Saif}
\affiliation{NASA Goddard Space Flight Center,  8800 Greenbelt Road, Greenbelt, Maryland 20771, United States}
\author{Linran Fan}
\email{linran.fan@utexas.edu}
\affiliation{Chandra Department of Electrical and Computer Engineering, The University of Texas at Austin, Austin, Texas 78758, United States}
\affiliation{James C. Wyant College of Optical Sciences, The University of Arizona, Tucson, Arizona 85721, United States}
\author{Saikat Guha}
\email{saikat@umd.edu}
\affiliation{Department of Electrical and Computer Engineering, The University of Maryland, College Park, Maryland 20742, United States}
\affiliation{James C. Wyant College of Optical Sciences, The University of Arizona, Tucson, Arizona 85721, United States}

\maketitle
\subsection{Abstract}
\textbf{
Achieving the ultimate Holevo limit of optical communication capacity requires a joint-detection receiver that makes a collective quantum measurement over multiple modulated symbols. Such superadditivity---a higher communication rate than is achievable by symbol-by-symbol optical detection---is a special case of the well-known nonlocality without entanglement and has yet to be demonstrated. 
In this article, we propose and demonstrate the design of a joint-detection receiver, the Green Machine, that can achieve superadditivity. 
We build this receiver, experimentally obtain the transition probability matrix induced by the codebook-receiver pair, and deduce that its capacity surpasses that of any symbol-by-symbol receiver in the photon-starved regime for binary-phase-shift-keying (BPSK)modulation. Our Green Machine receiver can also significantly reduce the transmitter peak power requirement compared with the pulse-position modulation (the conventional modulation format used for deep space laser communication). We further demonstrate that the self-referenced phase makes it resilient to phase noise, e.g., atmospheric turbulence or platform vibrations.
}

\section{Introduction}

\begin{figure*}[htbp]
\centering
\includegraphics[width=\linewidth]{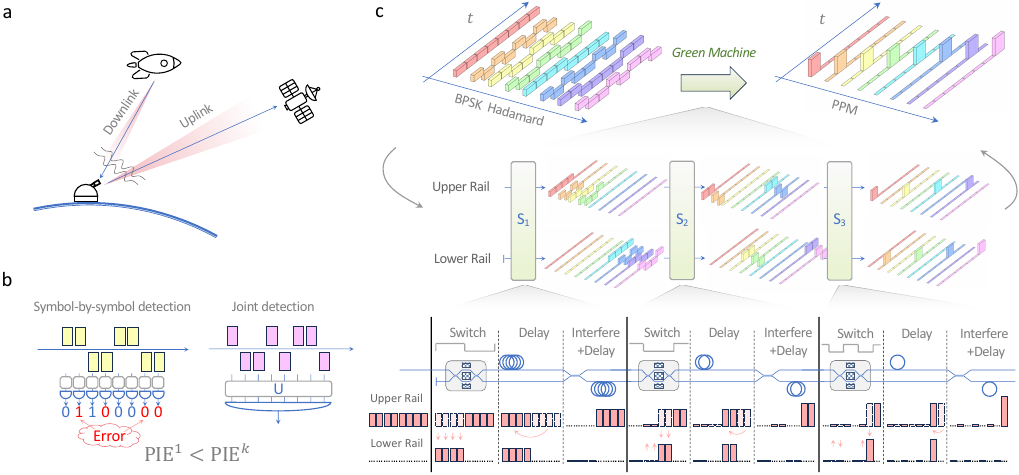}
\caption{\textbf{Schematics of superadditive communication with the Green Machine.} \textbf{a}. Uplink and downlink deep-space communication. The aperture sizes and range determine the loss at a given wavelength. Other channel impairments arise from atmospheric turbulence and imperfect pointing and tracking. \textbf{b}. In the photon-starved regime, symbol-wise decoding error dominates for symbol-by-symbol detection, which is why it can be outperformed by joint detection. \textbf{c}. The Green Machine joint-detection receiver applies a passive unitary Hadamard transform that maps BPSK Hadamard codewords to PPM codewords (each shown in a different color) prior to detection. The bottom schematic shows our compact design of the temporal Green Machine, which converts length-$N$ BPSK Hadamard codewords to $N$-ary PPM codewords using only $\log_2 N$ stages. $N=8$ (GM3) case is shown as an illustration.} 
\label{fig:Fig1} 
\end{figure*}

Higher communication capacity fuels the Information Age as the demand for faster information transfer continues to surge. Optical-frequency laser-light modulation is the leading choice for long-haul communication systems due to their far lower transmission losses compared with microwave signaling, both due to the advent of ultra-low-loss telecommunication equipment as well as far lower diffraction-induced spread for free-space communication. 

The fundamental limit of reliable optical communication rate, i.e., capacity, is governed by the laws of quantum mechanics and formulated as the Holevo-Schumacher-Westmoreland (HSW) theorem---colloquially known as the Holevo capacity~\cite{Schumacher1997,Holevo1998}. Holevo capacity is lower bounded by, and often exceeds, the Shannon capacity achievable by any receiver that uses symbol-by-symbol optical detection.
This gap becomes more prominent at the photon-starved regime where the received photon flux is low~\cite{Guha2012, Takeoka2014}. 
To close the gap, a joint-detection receiver applying a collective measurement across a codeword comprising multiple symbols is necessary. 
For a given encoding format, the fact that joint detection can achieve a bits-per-symbol rate that exceeds the highest attainable Shannon capacity by any symbol-by-symbol receiver is often termed superadditivity. This phenomenon is also a special case of the well-known quantum mechanical principle called Quantum Nonlocality without Entanglement~\cite{Bennett1999}.

Various operational scenarios drive a laser-based communication system into the photon-starved regime, where joint-detection receivers are advantageous. One example is deep-space optical communication (Fig.~\ref{fig:Fig1}a and Supplementary Note 2), where the transmitter laser's peak power is limited by satellite payload and power, and the aperture dimensions on both ends lead to a significant diffraction-limited loss over the extreme distances~\cite{hemmati2006deep,powell2013lasers}.

The relevant figure of merit to quantify the communication capacity in the photon-starved regime is the Photon Information Efficiency (PIE): the average number of bits (of information) reliably carried by each received photon. We denote ${\rm PIE}^k(\bar{n})=C^k(\bar{n})/k\bar{n}$, where $\bar{n}$ is the mean photon number per symbol pulse (modulation interval) at the receiver, and $C^k$ is the Shannon capacity over $k$ symbols for a given modulation format and an optical receiver. The superadditive capacity is achieved by a joint-detection receiver when ${\rm PIE}^k>{\rm PIE}^1$ with $k>1$ (Fig.~\ref{fig:Fig1}b). The Holevo limit of PIE can be interpreted as ${\rm PIE}^{k\rightarrow\infty}$.

In this paper, we report the realization of a joint-detection receiver called the optical Green Machine, which is capable of achieving superadditive capacity paired with the BPSK Hadamard codebook. BPSK modulation is capacity-optimal in the photon-starved regime with ${\rm PIE}^\infty(\bar n) = -\rm{\log_2}(\bar n) + 1$ bit per photon (bpp) up to two leading terms, approaching the Holevo-limited PIE of an unrestricted encoding~\cite{Banaszek2020}. However, the best-known symbol-by-symbol receivers for BPSK decoding can only achieve PIE of $2/\ln 2 \approx 2.89$ bpp. The Green Machine, first proposed as an $N$-port linear-optical interferometer~\cite{Guha2011}, 
transforms length-$N$ BPSK Hadamard codewords into $N$-ary Pulse Position Modulation (PPM) codewords prior to photon detection (Fig.~\ref{fig:Fig1}c). Therefore, the achievable capacity of the Green Machine follows the PPM performance in the photon-starved regime ${\rm PIE}^\infty(\bar n) = -\log_2(\bar n) - \log_2\ln (1/\bar n)$ bpp~\cite{Guha2012}, which can surpass the best symbol-by-symbol receivers for decoding BPSK (2.89 bpp). Moreover, it reduces the transmitter's peak power requirement by a factor of $N$ compared to PPM.

We build a fiber-based optical Green Machine first to realize the superadditive optical communication system~\cite{Guha2011}, and thereby demonstrate nonlocality without entanglement~\cite{Bennett1999} among a sequence of separated weak coherent states. The PIE achieved by decoding all the length-16 BPSK Hadamard codewords with our four-stage Green Machine outperforms ideal homodyne detection receivers after backing out losses. The demonstrated PIE also exceeds the best possible symbol-by-symbol receivers for distinguishing BPSK modulated signals, such as the Dolinar receiver, which can obtain the minimum error probability allowed by the laws of quantum mechanics~\cite{tsujino2011quantum,chen2012optical,rosati2017capacity,becerra2013experimental,dimario2018robust,cui2022quantum,burenkov2022experimental}.

We further verify that our Green Machine receiver is robust against channel phase noise by using the leading BPSK symbol as a common pilot phase reference across all codewords. This feature is inherent to the Hadamard encoding paired with the Green Machine, which makes its superadditive capacity even more prominent in noisy-phase environments, such as those caused by imperfect adaptive optics in a turbulent channel or platform vibrations. 

\begin{figure*}[htbp]
\centering
\includegraphics[width=\linewidth]{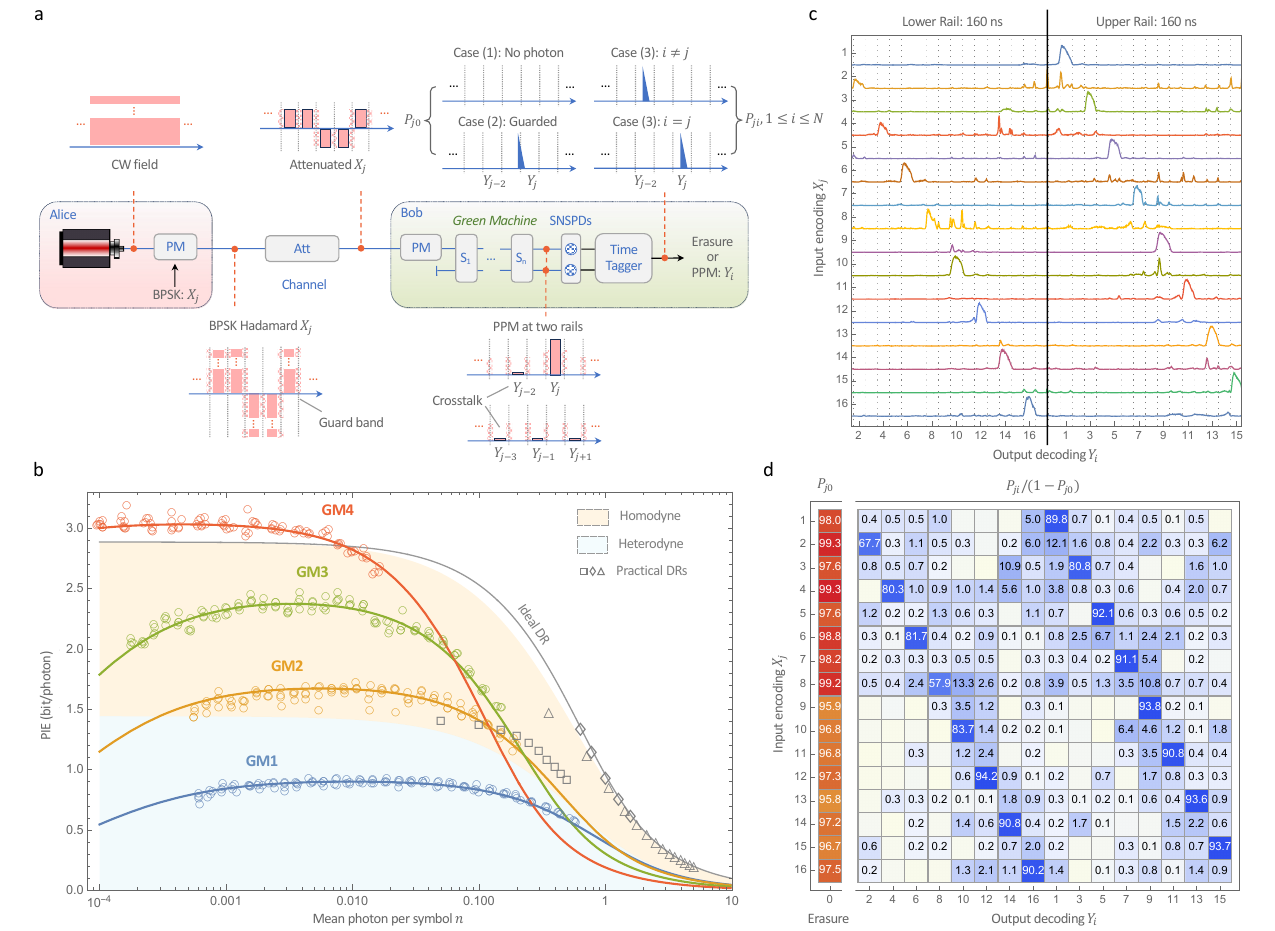}
\caption{\textbf{Four-stage Green Machine showing superadditivity with time-path multiplexing.} \textbf{a}. The diagram for GM experiments. PM (Alice): encoding phase modulator. Att: tunable optical attenuator. PM (Bob): phase modulators for correcting interferometers' phases in GMs (see Methods). \textbf{b}. PIEs for GMs with different numbers of stages and other state-of-the-art receivers. The experimentally obtained GM's PIEs are marked in colored circles with error bars smaller than the marker and thus omitted. The colored solid lines show the simulation results (GM4 has a lower dark count rate. See Methods). Loss-excluded practical Dolinar-type receivers (DR) are labeled in squares~\cite{cook2007optical}, triangles~\cite{dimario2018robust}, and diamonds~\cite{cui2022quantum} with the gray line showing the ideal Dolinar receiver's PIE. The light orange (blue) shade indicates the attainable soft PIE of homodyne (heterodyne) receivers~\cite{Banaszek2020}. \textbf{c}. Raw photon detection histograms of one set of GM4 experiments (different colors represent each of the $16$ codewords). \textbf{d}. The corresponding $N\times(N+1)$ transition probability matrix showing the erasure channel probability (warm colors) and normalized GM4-converted 16-ary PPM outputs (cold colors). Probabilities $<0.1\%$ are omitted.}
\label{fig:Fig2} 
\end{figure*}

\section{Results}
\subsection{Scalable Optical Green Machine} 

Realizing the Green Machine receiver based on its original proposal~\cite{Guha2011} would require deserializing $N$ pulses of the BPSK Hadamard code, applying lossless differential delays, followed by an $N$-port interferometer, which is challenging in practice.
Recently, a scalable design for the Green Machine was proposed that---via linear mixing of the time and path degrees of freedom---transforms temporally-encoded $N$-pulse-slot BPSK Hadamard codewords into $N$-PPM codewords, using only $\log_2 N$ stages, making the idea more feasible for an experimental implementation~\cite{TimRambothesis,banaszek2017structured,jachura2020scalable,Banaszek2020}. In this work, we further improve the scalable design using electro-optical switches to transform temporal modes into paths with practical and effective phase management. 

The central principle in designing a compact Green Machine is to apply simultaneous interferences among different temporal modes. As an example, we consider a three-stage Green Machine (GM3) that transforms a temporally-encoded BPSK Hadamard codeword consisting of eight symbols (Fig.~\ref{fig:Fig1}c). Each symbol is a weak coherent state with a temporal mode with duration $\tau$, and each Hadamard codeword contains eight symbols with equal amplitudes but binary-shifted phases. The complete codeword is received by the upper rail of the first stage (S1) of the Green Machine. 

In the first stage (S1), the four pairs of symbols that are $4\tau$ duration apart will interfere. These parallel processes are realized by switching the latter four symbols to the lower path and delaying the first four symbols in the upper path by $4\tau$. The switch is realized by the electro-optical modulator driven by square waves. Then, symbols separated by $4\tau$ in time can arrive at the beamsplitter simultaneously for interference. At the end of the first stage (S1), the lower path is delayed by $4\tau$ to avoid the degeneracy in temporal modes between the upper and lower paths. 
The second and third stages have the same structure, but different delays $2\tau$ and $\tau$. Therefore, symbols that are $2\tau$ and $\tau$ apart can interfere in the second and third stages, respectively. After the three stages, the input BPSK Hadamard codeword with eight symbols is transformed into an eight-ary PPM codeword, which can be decoded with single-photon detectors by reading the photon arrival time. The complete transformation process of all codewords is depicted in Fig.~\ref{fig:Fig1}c.

\begin{figure*}[htbp]
\centering
\includegraphics[width=\linewidth]{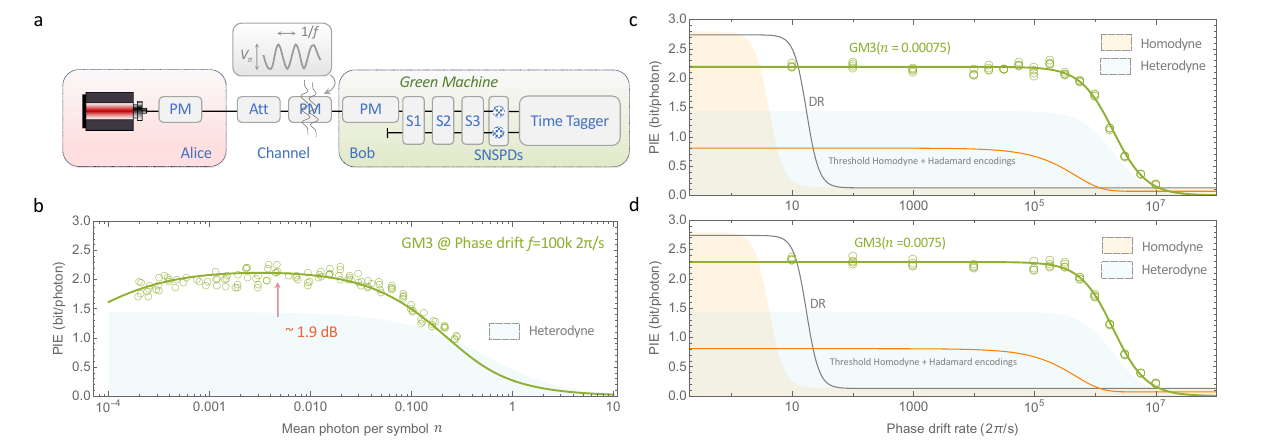}
\caption{\textbf{Demonstrating practical communication with a three-stage Green Machine showing immunity to channel phase noise.}\textbf{a}. The setup for testing the spectral response to channel phase drifts. The phase modulator in the channel scans the $2\pi$ phase at a single frequency $f$ to emulate channel phase drift. \textbf{b}. GM3's PIEs with phase drift at $10^{5}\times2\pi$/s. GM3's practical advantage is confirmed with a 1.9 dB higher PIE than the upper bound of the heterodyne receiver under fast channel phase drifting. \textbf{c}.  GM3's PIEs at $\bar{n}=7.5\times10^{-3}$ and \textbf{d}. at $\bar{n}=7.5\times10^{-4}$ with different channel phase drift rate. The Monte Carlo simulations of soft-information homodyne (light orange shades), Hadamard-structured homodyne (orange), and Dolinar (gray) receivers share the same conditions as the GM3 experiments under intractable phase drift.}
\label{fig:Fig3} 
\end{figure*}

\subsection{Superadditive Communication Demonstration}

We then implement our optical Green Machines for coherent optical communication with BPSK Hadamard encoding between the transmitter and receiver (Alice and Bob in Fig.~\ref{fig:Fig2}a). We test GM1, GM2, GM3, and GM4 for codeword length $N$=2, 4, 8, and 16, respectively. A variable attenuator in the channel is used to control the received mean photon number per symbol. 
At each channel loss, we repeat multiple experiments to benchmark the performances of the Green Machines with all $N$ codewords.

As shown in Fig.~\ref{fig:Fig2}a, there are three possible time-tagged photon detection events when Alice sends the $j^\mathrm{th}$ BPSK Hadamard codeword:
(1) no photon is detected across all $N$ time bins; 
(2) a photon is detected in the guard band; 
(3) a photon is detected in a time bin corresponding to the PPM codeword $Y_i,~1\leq i\leq N$, including the success case $i=j$ and error cases $i\neq j$. In the photon-starved regime with low crosstalk, the likelihood of more than one photon detected at multiple time bins is low. Thus, it is treated as invalid and discarded. For the $j^\mathrm{th}$ transmitted codeword in each run of the experiment, the erasure channel probability, composed of cases (1) and (2), is labeled as $P_{j0}$. For case (3), the corresponding success and error probabilities are collected into an array $P_{ji},~i=1,2,...,N$. Therefore, an $N$ by $N$+1 transition probability matrix is obtained by stacking all $N$ codewords. The average mean photon number per BPSK symbol is $\bar{n}=\sum_{j=1}^N \mathrm{ln}(P_{j0})/N^2$ by assuming Poisson distribution and equal priors for all codewords. Then, the ${\rm PIE}^N(\bar{n})$ is extracted by dividing the mutual information of the aforesaid $N\times$($N$+1) transition probability matrix by $N \bar{n}$.

The final PIE results are shown in Fig.~\ref{fig:Fig2}b. The acquired average crosstalks of GM1, GM2, GM3, and GM4 receivers are 1.0\%, 3.8\%, 7.0\%, and 11.3\%, while the highest attained PIE keeps increasing. The ${\rm PIE}^{16}$ achieved by the GM4 at $\bar{n}=0.0001$ to $\bar{n}=0.01$ outperforms the theoretical-best ${\rm PIE}^1$ achievable by any optical receiver employing symbol-by-symbol detection, including the state-of-the-art Dolinar receivers~\cite{cook2007optical,dimario2018robust,cui2022quantum} and the ideal homodyne detection receiver with soft-information decoding~\cite{Shapiro1985,Banaszek2020}. An exemplary photon detection histogram at $\bar{n}=0.00146\pm4\times10^{-6}$ with the associated transition probability matrix showing ${\rm PIE}^{16}=3.15\pm0.005$ are exhibited in Fig.~\ref{fig:Fig2}c and \ref{fig:Fig2}d. The gap between the PIEs of the GM4 and the ideal symbol-by-symbol detection receiver affirms quantum nonlocality without entanglement, which means that product quantum states can hold hidden information that is unattainable by symbol-by-symbol detection and classical post-processing~\cite{Bennett1999,pryde2005demonstrating}.

Additionally, the Green Machine receiver does not need any local oscillator (LO), as it utilizes the first symbol of the Hadamard codeword as a self-referenced phase reference to interfere with the following symbols~\cite{bartlett2007reference,Guha2011}. This feature gives the Green Machine robustness to counter channel-induced phase drift. To validate this benefit, we implement an artificial phase drift in the channel and test the performance of GM3 under phase drift (Fig.~\ref{fig:Fig3}a). We scan the mean photon numbers with a channel phase drift rate of 100 kilocycles per second (Fig.~\ref{fig:Fig3}b). Then, we select two $\bar{n}$ and scan the phase drift rate from $10$ cycles/s up to $10$ million cycles/s. The PIEs of GM3 attained with slow to fast phase drift are exhibited in Fig.~\ref{fig:Fig3}c and \ref{fig:Fig3}d, where the solid curves are fitted by an empirical low-pass band function $\frac{a}{1+(f/f_0)^{s}}$ (Fig.~\ref{fig:Fig3}c: $a=2.18$, $f_0=1.99$ MHz, and $s=1.68$; Fig.~\ref{fig:Fig3}d: $a=2.28$, $f_0=1.99$ MHz, and $s=1.58$). The performances of the ideal symbol-by-symbol homodyne and Dolinar receivers with no predictive phase tracking are fitted from the Monte Carlo simulations with identical data acquisition time and modulation bandwidth to the Green Machine experiments (total 2.5 million symbols, highlighted in Supplementary Note 5A). The heterodyne upper bound assumes perfect I/Q measurements over every symbol.

The above results support that our practical Green Machine receiver has nearly the same phase-drift tolerance as the theoretically optimal heterodyne detection receiver but obtains up to $1.9$ dB higher PIE. Conversely, the ideal Dolinar receiver and the shot-noise-limited homodyne detection receiver can not tolerate such rapid phase drift due to the lack of a receiver-internal self-phase reference mechanism~\cite{lami2023exact}. Therefore, the GM3 exhibits its superadditivity in the presence of strong phase drift as long as the rate of phase drift is significantly slower than the underlying BPSK modulation bandwidth. It could be contended that this advantage comes from using the Hadamard codewords. So, we also simulate the performance of the hard-threshold homodyne detection receiver on the same BPSK Hadamard codewords (solid orange lines in Fig.~\ref{fig:Fig3}c and \ref{fig:Fig3}d). Here, we see the tolerance to phase drift re-emerge, but three-quarters of the PIE is traded (compared with homodyne used on an unrestricted BPSK-modulated code) to gain phase stability due to the self-referenced phase of the BPSK Hadamard codewords. 

\section{Discussion}

In summary, we design and build a compact optical Green Machine receiver with superadditivity demonstrated at low mean photon numbers over a phase-drifting channel. The realized parallel processing of interfering the $N$-symbol Hadamard codewords into $N$-ary PPM codewords through $\log_2N$ stages, first proposed in Ref.~\cite{TimRambothesis,banaszek2017structured}, is crucial to the long-term scalability of this temporal Green Machine receiver. Our results represent a significant stepping stone towards attaining a reliable communication rate at the ultimate Holevo bound of channel capacity. So far, the communication system we have implemented only contains the optical realization of a pulsed laser transmitter, a fiber channel with loss and phase noise, and a receiver capable of realizing superadditive communication capacity. To demonstrate a fully functional communication system that can send messages reliably at a superadditive rate, one would need to incorporate an appropriate forward error correction code to correct the errors introduced by the superchannel induced by our Hadamard code Green Machine receiver combination—quantified by our measured transition probability matrix—such as the Reed-Solomon code or a low-density parity-check (LDPC) code, and an FPGA system that realizes the encoding and decoding operations.

Future directions include lowering crosstalk, adding more stages to the Green Machine, and investigating ways to introduce non-classical optical transforms, such as squeezing within the receiver. Eliminating or significantly reducing the losses within the receiver would be particularly important to achieve uncompromised superadditivity, i.e., superadditive capacity without backing out receiver-internal losses. The net loss of the current GM4 system is around $20$ dB, primarily introduced by commercial electro-optical modulators (EOMs). With improved coupling efficiency and device quality, the EOM loss could be made negligible~\cite{he2019low,li2023high}. If next-generation EOMs~\cite{wang2018integrated} coupled with ultra-low-loss integrated delay lines~\cite{chang2017heterogeneous} can reduce the loss to under $0.4$ dB per stage, scaling up to five stages will allow reaching the uncompromised superadditive ${\rm PIE}^{32}$ of $\approx3.0$ bpp, and retain immunity to phase noise in a realistic operational environment (see Supplementary Note 3). Another direction is to increase the modulation bandwidth towards GHz to compete with the state-of-the-art spectral efficiencies~\cite{kakarla2020one}, which in turn will reduce the lengths of the delay lines to sub-meter levels. This will help further reduce the size, loss, and design difficulty for a photonic-integrated Green Machine. The higher modulation bandwidth will also help further suppress the noise photons per modulated symbol. Finally, multiplexing single-photon detectors could help overcome the performance limit caused by a single detector's dead time. 

However, the pursuit of quantum receivers that achieve the Holevo capacity is far from being concluded with the Green Machine. Over the decades, many theoretical proposals have elucidated genuine joint-detection receivers capable of achieving superadditivity ~\cite{sasaki1998quantum,buck2000experimental,hastings2009superadditivity,czekaj2009purely,Guha2011,zhu2018superadditivity}, but practical solutions other than the Green Machine remain elusive. Moreover, there is strong theoretical evidence suggesting that joint-detection receivers describable by the semi-classical theory of photodetection---despite being superadditive-communication capable---cannot achieve the Holevo capacity~\cite{Chung2017,rosati2017capacity}. Instead, there are proposals of receivers that can achieve the Holevo capacity, e.g., (1) quantum polar code and successive cancellation receiver~\cite{Guha2012}, (2) the slicing receiver~\cite{Silva2013}, (3) codeword unambiguous state discrimination~\cite{Takeoka2013}, and (4) the sequential-decoding receiver~\cite{Wilde2012_Sequential} using the multimode collective Vacuum or Not Measurement~\cite{Oi2013}. The primary bottleneck in their practical realization is the need for high-fidelity or fault-tolerant quantum operations on optically-encoded information~\cite{Delaney2022}. Therefore, in the noisy intermediate-scale quantum (NISQ) era, the Green Machine receiver is undoubtedly a competitive near-term solution for photon-starved communication, which could afford a practical advantage over existing receivers.

Furthermore, the optical Green Machine we have built also has other applications, e.g., in (1) entanglement-assisted classical communication~\cite{Cox2023}, (2) reading information efficiently from a passively-encoded reflective classical memory~\cite{Guha2013}, and (3) loading the single-photon state across $N$ time bins into $\log_2 N$ quantum memories~\cite{Zheng2022}.

\section{Methods}

\subsection{Fiber-optical platform and control.} 

We implement polarization-maintaining fibers and components throughout the system at 1550 nm. The setup is mounted to an optical breadboard with a temperature sensor, an evenly distributed heater, and a PID controller loop to reduce phase noise caused by fiber motion and temperature fluctuation. The delay lines are 16 m for S1, 8 m for S2, 4 m for S3, and 2 m for S4. GM1 is demonstrated using S3; GM2 is composed of S2 followed by S3; GM3 is composed of S1, S2, and S3. For GM1, GM2, and GM3, the symbol time bin duration $\tau=20$  ns. GM4 is composed of S1, S2, S3, and S4 with $\tau=10$ ns.

Square waves control the switch modulators (EOSpace customized lithium niobate modules, 20GHz bandwidth) at near 6.25 MHz (S1), 6.25 MHz+$\pi/2$ phase (S2), 12.5 MHz+$\pi/2$ phase (S3) and 25.0 MHz+$\pi/2$ phase (S4). Phase encoding is achieved by loading a 6.25 MHz, a 12.5 MHz, a 25 MHz, and a 50 MHz square signal to PMs (Thorlabs, 10 GHz bandwidth). All signals are generated by arbitrary wave generators (Rigol DG4102 AWG and Rigol DG5102 AWG) capable of directly generating Vpp=10 V square waves.

\subsection{Interspersed active phase correction}. 

The crosstalk is only sensitive to the relative phase drift of two fiber links between the switch and the beamsplitter in each stage~\cite{jachura2020scalable}. The active phase correction is completed before Alice starts sending the encoded codewords. To correct the phase of each stage, we launch a bright probe laser light (Newfocus Telecom Laser 6428) and tap 1\% after each stage's beamsplitter. We scan the relative phase between symbols from 0 to $2\pi$ for $\log_2N$ times, each lasting a single period of 1 ms, by the PM (Bob) in front of the S1. The scan starts from the first stage. The phase control signal is first generated from an FPGA (RedPitaya STEMlab 125-14), amplified (SRS SIM911 BJT preamplifier), and sent as an amplitude modulation to the square wave signal generator (Rigol DG4102 AWG for S1, S2, and S3, and Rigol DG5102 AWG for S4) connected to the PM. The detector (Newport InGaAs Fiber-Optic Receiver 2011, tuned to 10 kHz bandwidth) reads the power envelope of the varying tapped light, and the output is collected by the same FPGA. The computer fits it by a sine function and triggers the FPGA output voltage to retain maximum or minimum interference. Immediately after, the phase control signal is set to this proper voltage to compensate for the phase difference. Similarly, in series, the phase drifts of all following stages are corrected one by one. The phase correction of each stage takes roughly 10 ms, which is constrained by the Ethernet/Serial port communication latency between the data acquisition units, the control units, and the computer. After phase correction, each relative phase can remain stable for roughly 100 ms in most scenarios. 

\subsection{Testing strategy} 
Alice, channel, and Bob are all controlled by one Python script running on the master computer. The script switches between phase correction and data acquisition. During the phase correction, the channel is switched to zero attenuation by a programmable optical attenuator (JAS Fitel HA9 Optical Attenuator). After phase correction, the channel is switched to a given high attenuation with artificial phase drift to emulate the actual channel. Alice starts encoding a given Hadamard codeword and keeps sending it for 50 ms, while Bob starts recording the photon detection events. If an error occurs during phase correction (e.g., fitting timed out), the data acquisition will be skipped to let the phase correction execute again.

\subsection{Data acquisition} 
The photon passing through the Green Machine is detected by Quantum Opus SNSPD (efficiency 85\% at 1550 nm, ideal dark count < 200 per second), tagged by Swabian Instruments TimeTagger Ultra (dead time 2 ns). The detection events of one round are saved to one individual local file, along with the synchronization clock ticking at 6.25 MHz. Since the synchronization clock is gated by an extra switch triggered by the computer, the actual communication time is usually less than 50 ms due to the inconsistency of Ethernet latency. During data processing, we manually set the guard band between PPM symbols to be 10 ns for GM1, GM2, and GM3 tests, and 8 ns for GM4 tests. The guarded symbol lengths are 10 ns for GM1, GM2, and GM3 and 2 ns for GM4.

During the experiment, the GM1, GM2, and GM3 data are continuously collected over a few hours, including sweeping the channel loss and dialing the codewords. The GM4 data are continuously collected over tens of hours as the number of codewords is doubled, more data sets are collected, and the phase correction error occurs more. For GM4, each set of experiments dials through all 16 codewords twice so that the data can be combined to reduce statistical uncertainty. The bottleneck of data acquisition time is saving large files returned from the Time Tagger.

The estimated noise photon rate, including dark counts and ambient light, for GM1, GM2, and GM3 is around $4\times10^{-5}$ per 10 ns guarded symbol over two SNSPDs. For GM4, since its guarded-symbol length is shorter and the dead time after guard-band count affects the register of possible noise photons, we anticipate GM4's effective noise photon is reduced to $2\times10^{-6}$ per symbol according to our model, resulting in a slower drop of GM4's PIE at $\bar{n}<0.001$ side shown in Fig.~\ref{fig:Fig2}b. The data processing and the generation of data-related figures in this work are done using Mathematica, Version 14.2,(2024)~\cite{Mathematica}

For more details, please see Supplementary Note 5.

\section{Data availability}

The raw detection data of one run of the GM3 that was generated in this study, and the corresponding data processing code have been deposited in the Zenodo available at URL: \url{https://doi.org/10.5281/zenodo.10215208}.

\section{References}
\bibliography{Ref}

\section{Acknowledgments}
The authors acknowledge the support from NASA (grant number 80NSSC22K1030) and the NSF ERC Center for Quantum Networks (grant number EEC-1941583). CC and JP thank Allison Rubenok, Khanh Kieu, and Lam Nguyen for their assistance with PM-fiber splicing, thank Rohan Bali for assistance on FPGA, and thank Yiyun Wu and Samar Choura for helping with the thermally stabilized optical breadboard and 3D-printed parts. SG thanks Konrad Banaszek for comments on the performance of homodyne and heterodyne detection receivers with soft information processing.

\section{Author Contributions Statement}
All authors conceived the idea and wrote the manuscript together. CC, JP, and LF designed the experiment. CC and JP conducted the experiments and analyzed the data. CC, JP, and SG derived the theoretical model. LF, BS, and SG supervised the work.

\section{Competing Interests Statement}
The authors declare no competing interests.

\clearpage
\newpage

\title{Supplementary Information:\\ Superadditive Communication with the Green Machine as a Practical Demonstration of Nonlocality without Entanglement}

\maketitle

\onecolumngrid

\large

\setcounter{figure}{0}
\renewcommand{\theequation}{S\thesection\arabic{equation}} 
\renewcommand{\figurename}{\textbf{Supplementary Figure}}
\renewcommand{\thefigure}{\textbf{\arabic{figure}}}
\renewcommand{\tablename}{\textbf{Supplementary Table}}
\renewcommand{\thetable}{\textbf{\arabic{table}}}

\newpage

\section{Supplementary Note 1: Historical Green Machine}
Despite being commonly recognized as fruity drinks or children's toys, the idiom "Green Machine" has been associated with the Hadamard Codebook for over 60 years. One Richard R. Green of NASA's Jet Propulsion Laboratory, when pursuing his Ph.D. at Caltech, proposed a "Serial Orthogonal Decoder"\cite{Green1966} for a specific set of block codewords defined by sequentially-tensored Hadamard matrices. A traditional decoder for the book would require $2N$ addition or subtraction actions for the $N$-length code, which at the time was a high overhead for computing devices. By recognizing that the code was able to be transformed through a series of permutation operations, Green devised a design for a scalable "machine" capable of decoding the codebook with $\log_2N$ identical stages for an arbitrary $N$-length Hadamard codebook. The results were promising, showing that a 7-stage "machine" would be capable of achieving >35 kbps transmission.

Green's idea was put to use only a few years later for the Mariner-9 mission, the first device to orbit another planet. Mission designers sought to transmit gray-scale images of the Martian surface back to Earth using binary 6-tuples representing the 64-bit pixel values. With a channel error rate of $5\%$, each pixel error rate would be $26\%$. This failure rate was far too high for such a critical mission, especially considering the intention to transmit the data as a live feed and orient the camera in response to what was seen. Due to power limitations, the mission allowed a forward-error-correction code to embed the 6-tuples into 30 channel uses. The mission designers finally chose to use the first-order Reed-Solomon (Hadamard) code with Green's circuit design and internally dubbed it "The Green Machine". 

The current Green Machine receiver design stems directly from co-author Saikat Guha's 2011 paper~\cite{Guha2011} on using Joint-Detection Receivers (JDR) to approach the Holevo capacity over classic and quantum communication channels. The idea for the compact time-multiplexed optical  Green Machine was first considered and attempted experimentally by Timothy Rambo at Northwestern University, Illinois, USA, who was able to realize a single-stage device experimentally~\cite{TimRambothesis} and independently proposed by Konrad Banaszek and Michał Jachura~\cite{banaszek2017structured,jachura2020scalable} at University of Warsaw, Warsaw, Poland, at almost the same time. This work, in a similar vein, brings to fruition the first realization of a compact multi-stage Green Machine JDR for superadditive classical-quantum optical communications.

\newpage

\section{Supplementary Note 2: Deep-Space Communications and the Advantageous Range of the Green Machine}

To discuss the scenario where our optical Green Machine receiver can show a practical advantage, it is necessary to know the received mean photon number and the equivalent distance in the current deep-space communication testing platforms. Here, we compare the power budget, i.e., the mean received photon number, of various communication systems deployed in practice and then show the regime where our current GM design can have advantages if applied in these missions.

A coherent signal with a fundamental Hermite-Gaussian mode propagating through a free space link will experience diffraction-induced losses due to the limited aperture size of the receiving system. The power received for aperture-limited free space communication link can be compactly written as
\begin{equation}
    P_r = P_w \left(\frac{\pi D_t}{\lambda}\right)^2 \left(\frac{\pi D_r}{\lambda}\right)^2 \left(\frac{\lambda}{4 \pi R}\right)^2 \eta,
\end{equation}
with parameters defined in the \textbf{Supplementary Table 1}:
\begin{table}[h]
    \centering
    \begin{tabular}{l|r}
        $P_r$ ~& Received Power \\
        $P_w$ ~& Transmitted Power  \\
        $D_t$ ~& ~~~Diameter of transmitter  \\
        $D_r$ ~& Diameter of receiver  \\
        $R$ ~& Distance traveled  \\
        $\lambda$ ~& Carrier wavelength \\
        $\eta$ ~& System efficiency 
    \end{tabular}
    \caption{Definitions of parameters for estimating power budget in deep-space laser communication platforms.}
    \label{tab:my_label}
\end{table}

For a baseline comparison, we choose to investigate the Lunar Lasercom Demonstration (LLCD) and the Deep Space Optical Communication (DSOC) package operating at Lunar, Martian, and Psyche distances. The relevant parameters for each system are then listed in \textbf{Supplementary Table 2} (uplink) and \textbf{Supplementary Table 3} (downlink).

\begin{table}[]
    \centering
\begin{tabular}{|| c | c c c c||} 
 \hline
 params & LLCD & DSOC (L) & DSOC (M) & DSOC (P) \\ [0.5ex] 
 \hline\hline
 $P_r (W) $ & $3.84\times10^{-10}$ & $1.15\times10^{-4}$ & $2.60\times10^{-10}$ & $2.11\times10^{-10}$ \\ \hline
 $P_w (W)$ & 10 & 5000 & 5000 & 5000 \\ 
 \hline
 $D_t (m)$ & 0.15 & 1 & 1 & 1 \\
 \hline
 $D_r (m) $ & 0.1 & 0.22 & 0.22 & 0.22 \\
 \hline
 $R (m)$ & $388.1\times10^6$ & $388.1\times10^6$ & $225\times10^9$ & $250\times10^9$ \\
 \hline
 $\lambda (um)$ & 1.55 & 1.064 & 1.064 & 1.064 \\
 \hline
 $\eta$ & 0.1* & 0.1* & 0.1* & 0.1* \\ [1ex] 
 \hline
\end{tabular}
\caption{Uplink parameters of current deep-space communication platforms operated by NASA.}
\end{table}
\begin{table}[]
    \centering
\begin{tabular}{|| c | c c c c||} 
 \hline
 params & LLCD & DSOC (L) & DSOC (M) & DSOC (P) \\ [0.5ex] 
 \hline\hline
 $P_r (W) $ & $1.364\times10^{-10}$ & $8.25\times10^{-9}$ & $2.45\times10^{-12}$ & $1.99\times10^{-12}$\\ \hline
 $P_w (W)$ & 0.5 & 4 & 4 & 4 \\ 
 \hline
 $D_t (m)$ & 0.1 & 0.22 & 0.22 & 0.22 \\
 \hline
 $D_r (m) $ & 0.4 & 5 & 5 & 5 \\
 \hline
 $R (m)$ & $388.1\times10^6$ & $388.1\times10^6$ & $225\times10^9$ & $250\times10^9$ \\
 \hline
 $\lambda (um)$ & 1.55 & 1.55 & 1.55 & 1.55 \\
 \hline
 $\eta$ & 0.1* & 0.1* & 0.1* & 0.1* \\ [1ex] 
 \hline
\end{tabular}
\caption{Downlink parameters of current deep-space communication platforms operated by NASA.}
\end{table}

The distances shown for Martian and Lunar communication systems are the average values, while for the Psyche, it is listed as the distance for the 2024 nearest approach. System efficiency is given as an estimate, about 0.1 (marked with *). Both the current DSOC and previous LLCD missions use pulse position modulation for communications. Our following simulations use the pulse duration of $2ns$. The average number of photons is presented as
\begin{equation}
    \bar{n}=P_w \times \frac{\lambda\tau}{h c}
\end{equation}

\begin{figure}[htbp]
    \centering
    \includegraphics[width=3.5in]{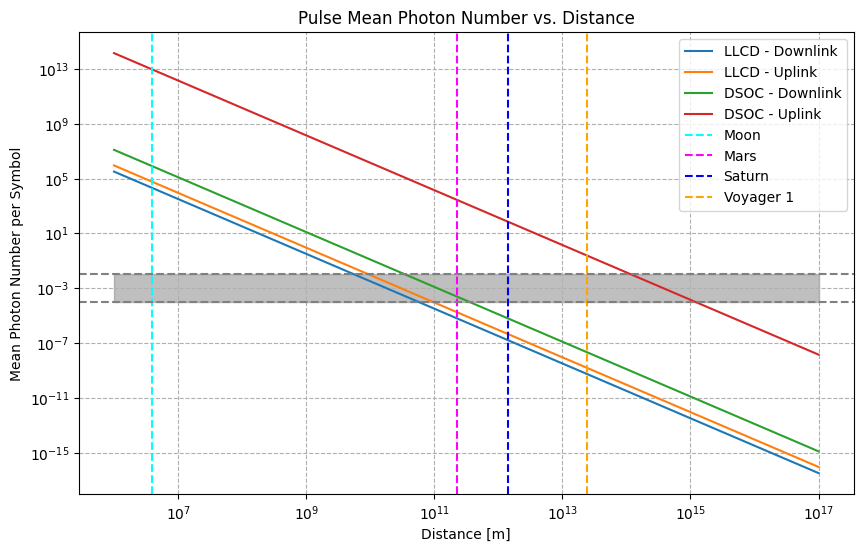}
    \vspace{-4pt}
    \caption{Simulations of the mean photon number per pulse budget in current deep-space communication platforms operated by NASA. Solid lines indicate the received mean photon per pulse for various deep-space communication systems, including LLCD downlink (blue) and uplink (orange), DSOC mission downlink (green), and uplink (red). Dashed vertical lines indicate the distance of various celestial bodies of interest. The gray highlighted box indicates the region of mean photon number that our current GM4 is demonstrating superadditive capacity. Below the gray box, the GM with more stages could show superadditivity as it can compress a longer length of BPSK Hadamard codewords where each symbol can have fewer photons, as shown in \textbf{Supplementary Figure}~\ref{fig:SuppS1}\textbf{a}.}
    \label{fig:SuppS0}
\end{figure}

Currently, one limiting factor of the pulsed-laser power budget is the load/cost of the customized-high-power Q-switch laser and the gain media it equips. In principle, a customized CW laser could have more average power output, a smaller size, and a lower weight than these pulsed lasers. If we managed to build a GM working at the same modulation rates but using length-$N$ BPSK Hadamard codewords encoded over a CW laser beam, the final converted $N$-PPM pulses would attain more photons at higher capacities. We anticipate the GM architecture will be the next paradigm of deep-space communication, especially when the spacecrafts are reaching further, or the human successfully establishes a colony on Mars that requires high-quality, high-capacity, and constant communication. Therefore, it is now necessary to visualize the mean photon for a received PPM pulse for a given communication system as a function of the distance separated. \textbf{Supplementary Figure}~\ref{fig:SuppS0} shows the power budget of both LLCD and DSOC platforms' uplink and downlink, and the sweet spot that our current GM4 experiment has shown advantages. For reference, the average distances of the Moon, Mars, Saturn, and Voyager 1 current distance are shown.

\section{Supplementary Note 3: Theoretical Backgrounds in Brief}

\subsection{A. BPSK, PPM, and Hadamard Codewords in Optical Communication}

The quantized electromagnetic field from an ideal CW laser in a single spatio-temporal mode can be described as a coherent state by the complex amplitude $\alpha$.
\begin{equation}
   \ket{\alpha}=e^{-|\alpha|^2/2}\sum_{n}\frac{\alpha^n}{\sqrt{n!}}\ket{n}
\end{equation}

Binary Phase-Shifted Keying (BPSK) is a common modulation format used across digital communications. In such an optical communication system, the modulation embeds information into coherent state symbols picked from the alphabet $\mathcal{A}_{\rm BPSK}=\{\ket{\alpha},\ket{-\alpha}\}$. For this alphabet, the transmitter generates one of the two coherent states lasting for a duration of $\tau$ seconds. After transmitting through the channel, the signal is taken by an optical receiver that discriminates between the two states.

Based on Shannon's information theory and the fundamental quantum properties of the coherent states, the symbol-by-symbol encoding and decoding of the BPSK signals are bounded to the Photon Information Efficiency (PIE) of $2\log_2e\approx2.89$ bits per photon (bpp). Grouping the symbols into codewords allows forward error correction, which is essential in modern communication systems. One iconic format for grouping up BPSK symbols is the Hadamard code, also known as the Reed-Solomon code, which is constructed from the Hadamard–Rademacher–Walsh transform matrix. The eight-symbol standard Hadamard BPSK codewords carried by eight coherent states are
\begin{equation}
\begin{aligned}
&\ket{X_1}=\ket{\alpha}\otimes\ket{\alpha}\otimes\ket{\alpha}\otimes\ket{\alpha}\otimes\ket{\alpha}\otimes\ket{\alpha}\otimes\ket{\alpha}\otimes\ket{\alpha}\\
&\ket{X_2}=\ket{\alpha}\otimes\ket{-\alpha}\otimes\ket{\alpha}\otimes\ket{-\alpha}\otimes\ket{\alpha}\otimes\ket{-\alpha}\otimes\ket{\alpha}\otimes\ket{-\alpha}\\
&\ket{X_3}=\ket{\alpha}\otimes\ket{\alpha}\otimes\ket{-\alpha}\otimes\ket{-\alpha}\otimes\ket{\alpha}\otimes\ket{\alpha}\otimes\ket{-\alpha}\otimes\ket{-\alpha}\\
&\ket{X_4}=\ket{\alpha}\otimes\ket{-\alpha}\otimes\ket{-\alpha}\otimes\ket{\alpha}\otimes\ket{\alpha}\otimes\ket{-\alpha}\otimes\ket{-\alpha}\otimes\ket{\alpha}\\
&\ket{X_5}=\ket{\alpha}\otimes\ket{\alpha}\otimes\ket{\alpha}\otimes\ket{\alpha}\otimes\ket{-\alpha}\otimes\ket{-\alpha}\otimes\ket{-\alpha}\otimes\ket{-\alpha}\\
&\ket{X_6}=\ket{\alpha}\otimes\ket{-\alpha}\otimes\ket{\alpha}\otimes\ket{-\alpha}\otimes\ket{-\alpha}\otimes\ket{\alpha}\otimes\ket{-\alpha}\otimes\ket{\alpha}\\
&\ket{X_7}=\ket{\alpha}\otimes\ket{\alpha}\otimes\ket{-\alpha}\otimes\ket{-\alpha}\otimes\ket{-\alpha}\otimes\ket{-\alpha}\otimes\ket{\alpha}\otimes\ket{\alpha}\\
&\ket{X_8}=\ket{\alpha}\otimes\ket{-\alpha}\otimes\ket{-\alpha}\otimes\ket{\alpha}\otimes\ket{-\alpha}\otimes\ket{\alpha}\otimes\ket{\alpha}\otimes\ket{-\alpha}\\
\end{aligned}
\end{equation}

Pulse-position modulation (PPM) is another widely used modulation format in photon-starved regimes, such as deep-space communication. Combined with direct photon detection, the PPM can reach higher PIE than BPSK with classical detection but requires higher peak power, which poses a challenge to the transmitter. With the same mean photon numbers per codeword as BPSK, the eight-ary PPM codewords are

\begin{equation}
\begin{aligned}
&\ket{Y_1}=\ket{\sqrt{8}\alpha}\otimes\ket{0}\otimes\ket{0}\otimes\ket{0}\otimes\ket{0}\otimes\ket{0}\otimes\ket{0}\otimes\ket{0}\\
&\ket{Y_2}=\ket{0}\otimes\ket{\sqrt{8}\alpha}\otimes\ket{0}\otimes\ket{0}\otimes\ket{0}\otimes\ket{0}\otimes\ket{0}\otimes\ket{0}\\
&\ket{Y_3}=\ket{0}\otimes\ket{0}\otimes\ket{\sqrt{8}\alpha}\otimes\ket{0}\otimes\ket{0}\otimes\ket{0}\otimes\ket{0}\otimes\ket{0}\\
&\ket{Y_4}=\ket{0}\otimes\ket{0}\otimes\ket{0}\otimes\ket{\sqrt{8}\alpha}\otimes\ket{0}\otimes\ket{0}\otimes\ket{0}\otimes\ket{0}\\
&\ket{Y_5}=\ket{0}\otimes\ket{0}\otimes\ket{0}\otimes\ket{0}\otimes\ket{\sqrt{8}\alpha}\otimes\ket{0}\otimes\ket{0}\otimes\ket{0}\\
&\ket{Y_6}=\ket{0}\otimes\ket{0}\otimes\ket{0}\otimes\ket{0}\otimes\ket{0}\otimes\ket{\sqrt{8}\alpha}\otimes\ket{0}\otimes\ket{0}\\
&\ket{Y_7}=\ket{0}\otimes\ket{0}\otimes\ket{0}\otimes\ket{0}\otimes\ket{0}\otimes\ket{0}\otimes\ket{\sqrt{8}\alpha}\otimes\ket{0}\\
&\ket{Y_8}=\ket{0}\otimes\ket{0}\otimes\ket{0}\otimes\ket{0}\otimes\ket{0}\otimes\ket{0}\otimes\ket{0}\otimes\ket{\sqrt{8}\alpha}\\
\end{aligned}
\end{equation}

\subsection{B. Receivers and Their Capacities}

Based on Shannon's information theory, the classical channel capacity defines the amount of information able to be transmitted in a single channel use (codeword sent).
\begin{equation}
   C(\bar{n})=\max_{p_X}\biggl[I(X;Y)\biggr],
\end{equation}
where $I(X;Y)$ is the classic mutual information. In our case of optical communication,  the transmitter encodes $X$ into a light field, and the receiver's outputs are $Y$ with $\bar{n}$ photons received per channel per code on average. Then, the PIE is defined as PIE=$C(\bar{n})/\bar{n}$. Here, output $Y$ can have one more choice as an erasure channel for no-photon-arriving cases, which is necessary, especially when $\bar{n}$ is small.

In photon-starved communications, the Holevo PIE, a.k.a. the quantum-mechanically-allowed highest possible PIE $\approx \log_2\frac{1}{\bar{n}}+\log_2e$ bpp without specifying the transmitter or receiver, is attainable using BPSK encoding, while any symbol-by-symbol detection receiver for decoding BPSK can not reach to this limit. In this work, we discuss three types of practical symbol-by-symbol detection receivers that are widely known for decoding BPSK. (1) The homodyne receiver has the PIE $\approx\frac{1}{2\bar{n}}\log_2(1+4\bar{n})$ bpp with the cost of needing a robust high-power local oscillator with perfect phase reference to the source. (2) The heterodyne receiver does not rely on a rigorous phase reference due to an extra quadrature to correct the phase during the measurement. But its performance drops to PIE $\approx\frac{1}{2\bar{n}}\log_2(1+2\bar{n})$ bpp. (3) The Dolinar receiver~\cite{Dolinar1973}, which uses an adaptive dynamic local oscillator combined with direct detection, can approach the Helstrom-bound-limited bit error rate, a.k.a the quantum-mechanically-allowed minimum bit error rate, in decoding BPSK signals. 
\begin{equation}
    P_e(\bar{n})=\frac{1}{2}(1-\sqrt{1-e^{-\bar{n}}})
\end{equation}
However, its PIE $=\frac{1}{\bar{n}}\left[1-P_e(\bar{n})\log_2\frac{1}{P_e(\bar{n})}-(1-P_e(\bar{n}))\log_2\frac{1}{1-P_e(\bar{n})}\right]$ bpp is still bounded to $2\log_2e\approx2.89$  bpp in photon-starved communications.

When considering the photon-starved regime, the PPM encoding performs relatively well in terms of PIE, although it cannot theoretically approach the Holevo PIE~\cite{Banaszek2020}. An important point of distinction is that PPM with direct detection does not rely on any local oscillator or phase reference since the free phase parameters of each codeword are eliminated by direct detection on power or photon numbers. The Green Machine we build is to transform the $N$-symbol Hadamard codewords into $N$-PPM codewords within $\log_2N$ stages so that it not only cuts the need for high-peak-power laser but also enjoys the PPM's advantages on a higher PIE and immunity to channel phase drift, which we will show later. The PIE of the $N$-symbol Hadamard codebook with the GM receiver has the same capacity as the $N$-PPM with direct photon detection.
\begin{equation}
    {\rm PIE}^{N}_{\rm GM/PPM}=\frac{1-e^{-N\bar{n}}}{N\bar{n}}\log_2N
\end{equation}
For more details, we suggest the review article~\cite{Banaszek2020} by Konrad Banaszek. 

\subsection{C. Superadditive communication and nonlocality without entanglement}

Superadditive communication is achieved when the PIE of a joint-detection receiver is higher than that of any symbol-by-symbol detection with any electronic postprocessing. It means adding one symbol to the codeword gains information higher than the information contained in that symbol individually. If no hidden information is embedded in the concatenation of symbols, adding more symbols will only linearly increase the information, a.k.a. an additive channel. That is to say, the existence of superadditive communication means that applying proper joint detection can extract more information from a product state than any symbol-by-symbol detection strategy.  

From the quantum information perspective, it also means this quantum system, even without entanglement, can show some relation among the subsystems. Since these subsystems are generated in a time series, they have never talked to each other before arriving at the measurement apparatus. This infers that this type of relation is a facet of nonlocality that is not necessarily bound to quantum entanglement, which is first introduced in Charles Bennet's paper~\cite{Bennett1999}: {\em ``an orthogonal set of product states (e.g., of two three-state particles) that nevertheless cannot be reliably distinguished by a pair of separated observers ignorant of which of the states has been presented to them, even if the observers are allowed any sequence of local operations and classical communication between the separate observers.''}. 

\begin{figure}[htbp]
    \centering
    \includegraphics[width=3.5in]{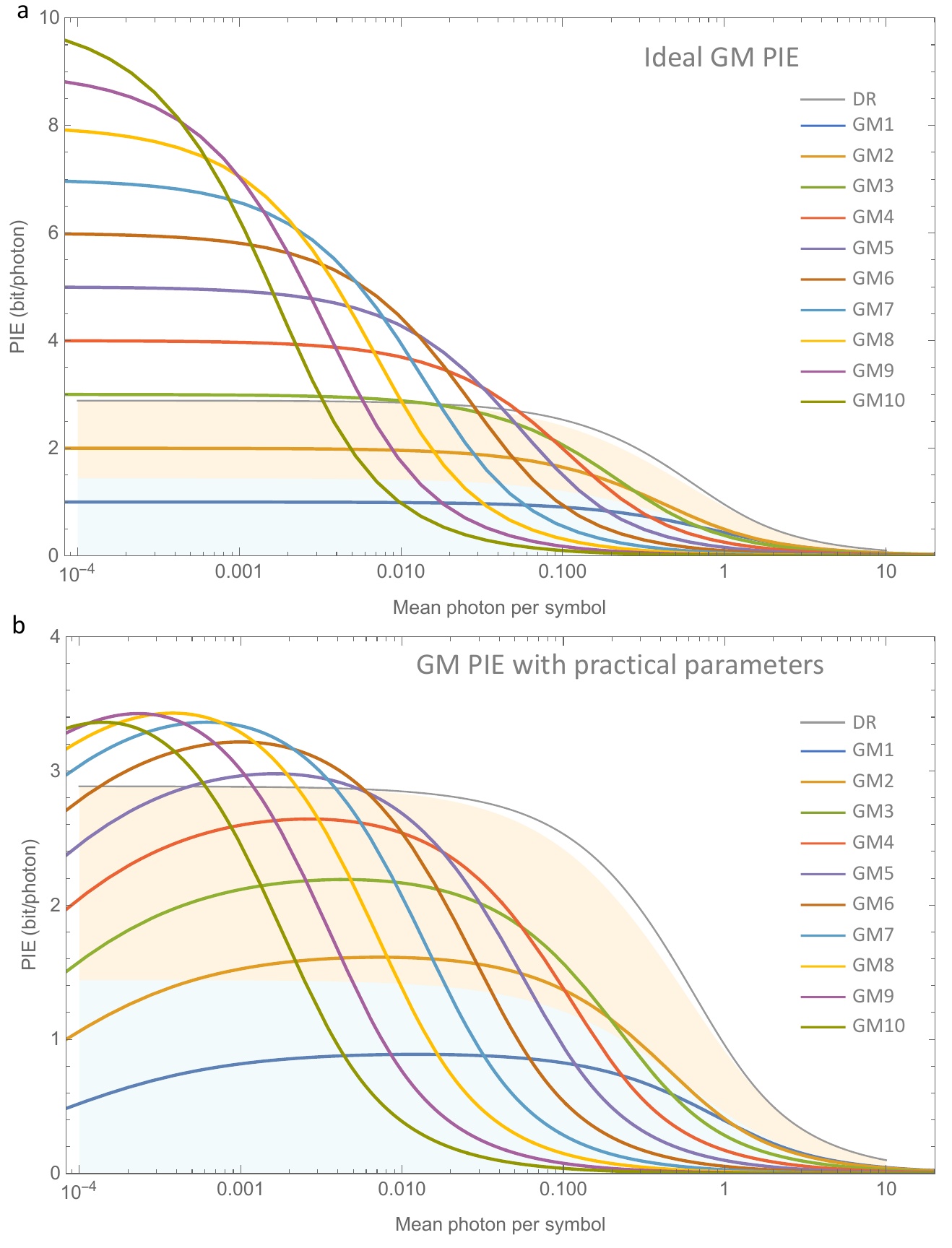}
    \vspace{-4pt}
    \caption{Simulations. Orange (blue) shade is for the homodyne (heterodyne) receiver. The gray line is for the Dolinar receiver (DR). \textbf{a}, PIEs of ideal multi-stage Green Machines. \textbf{b}, PIEs of practical multi-stage Green Machines with 0.4 dB loss per stage and $4\times10^{-5}$ noise photon per symbol (considering ambient light and fiber leakage of the fiber components and the SNSPD's intrinsic dark count). GM8 is almost the best performance under these circumstances.}
    \label{fig:SuppS1}
\end{figure}

Our current experiment has demonstrated superadditive communication and nonlocality without entanglement by showing that a four-stage Green Machine (GM4) receiver decoding length-16 Hadamard BPSK secures a higher PIE than any symbol-by-symbol detections in the photon-starving regime. In the current interpretation, we exclude the net loss of all the receivers. This is a practical distinction for this claim since heterodyne and homodyne receivers usually have much lower losses than the current GM4 setup. However, the current loss of GM4 does not come from any fundamental physical limit, which means it could be improved with better devices, especially the electro-optical modulators. The feasibility of developing low-loss Green Machine receivers will be an important question in forthcoming extensions of this work. \textbf{Supplementary Figure}~\ref{fig:SuppS1} shows the PIE simulations of Green Machines with different numbers of stages and other competitors under ideal conditions and with realistic imperfections.

\newpage

\section{Supplementary Note 4: Phase Symbol, Channel Phase Noise, and Monte Carlo Simulations}

The first symbol of Hadamard BPSK code adds no distinguishability to the codewords. Thus, it can be generated either at the receiver or transmitter sides. In real-world applications where phase reference is hard to track, such as deep-space communication, this phase symbol applied at the transmitter side will experience channel properties together with the rest of the symbols. The timescale of the codeword is so short that the channel phase remains nearly the same. Therefore, having this phase symbol will make the communication system robust to channel phase noise. 

To illustrate this advantage over other symbol-by-symbol detection receivers, we deploy the Monte Carlo simulations to predict the performances of ideal homodyne, collective-threshold homodyne, and ideal Dolinar receivers under different channel phase drift rates in a certain amount of time ($50$ ms). 

(I) The ideal symbol-by-symbol homodyne/heterodyne receiver decodes each symbol as a real number based on the output of homodyne/heterodyne detection, where more Shannon information can be extracted by including forward-error correction and other post-processing~\cite{Banaszek2020}. 

(II) The threshold homodyne receiver directly decodes each symbol of BPSK based on the sign of homodyne detection, yielding a "0" if the sign is positive or a "1" if negative. The threshold homodyne receiver is more flexible in practical implementation as it fits short codewords with digitized collective post-processing, including the length-eight Hadamard code used in our work. Therefore, we anticipate it to be a direct competitor toward the Green Machine.

(III) The ideal Dolinar receiver in the simulation applies an immediate feedforward strategy~\cite{assalini2011revisiting} to change the phase of the local oscillator according to photon detection history within each symbol. The ideal Dolinar receiver reaches the lowest possible error rate (Helstrom bound) in decoding the BPSK symbol. 

\begin{figure}[htbp]
    \centering
    \includegraphics[width=4in]{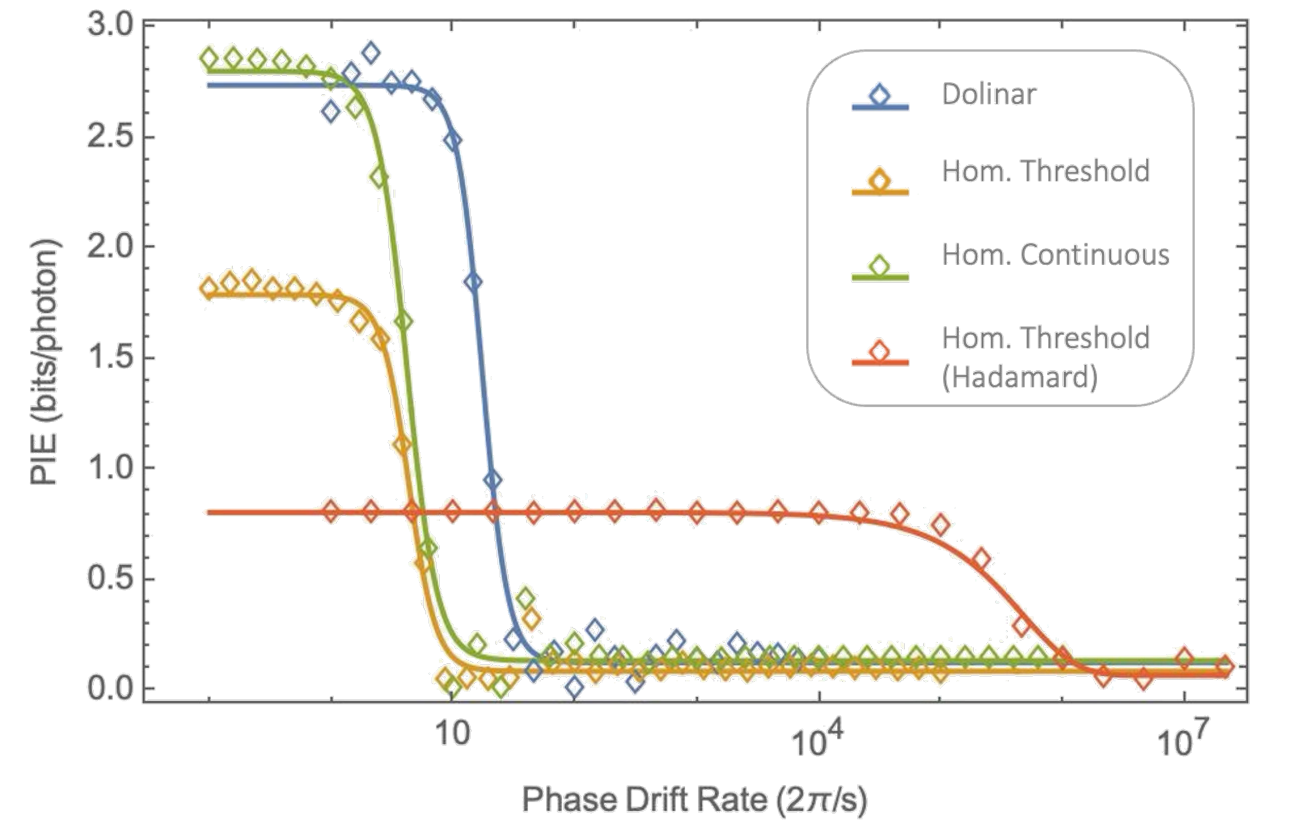}
    \caption{Monte Carlo simulation results for PIEs w.r.t different phase drift rates. Each point contains 2.5 million trials.}
    \label{fig:SuppS2}
\end{figure}

The Monte Carlo results shown in \textbf{Supplementary Figure}~\ref{fig:SuppS2} indicate that the PIEs of Dolinar and homodyne receivers that need an extra phase reference to generate a local oscillator start dropping at 10Hz-level phase noise. Using the same eight-symbol Hadamard encoding and threshold homodyne receiver for decoding will provide more robustness to phase noise but reduce the PIE. On the contrary, the heterodyne receiver can leverage the second quadrature information to correct the phase drift but suffers a reduced PIE. Moreover, if the phase drift is as fast as the modulation bandwidth, the signal arrived at the receiver will be in a mixed state; thus, none of the receivers can extract information from it. To not compromise our claim of practical advantage, we aggressively assume that the upper bound of an ideal heterodyne receiver can retain its PIE until the phase drift approaches the modulation bandwidth with the best possible strategy of phase correction leveraging the extra quadrature. However, in practice, the PIE of the heterodyne receiver will drop at a much slower phase-drift rate due to other practical imperfections. 

In a typical free-space laser communications downlink, the phase drift caused by the Earth's atmospheric turbulence is usually below 100$\pi/s$, corresponding to the Greenwood frequency~\cite{greenwood1977bandwidth}. The standard adaptive optical system and deformable mirror used in optical telescopes usually have an update rate of 1 kHz (1 frame per ms). Therefore, the maximum Hadamard codeword duration with phase drift immunity is around $1$ms, which corresponds to $2^{16}$ time bins each $10$ns long (can be converted into PPM of order $2^{16}$ by a $16$-stage GM). By reducing the bin length with high-bandwidth electronics, the order can be further increased. It is within current technological reach to build a Green Machine for order $2^{19}$ and be compatible with the state-of-the-art PPM experiments, albeit with $2^{19} \sim 500$K-fold lower peak transmitted power~\cite{rielander2023esa,guo2023record}.

\newpage

\section{Supplementary Note 5: Experimental Setup and Calibration}

\begin{figure}[htbp]
    \centering
    \includegraphics[width=5in]{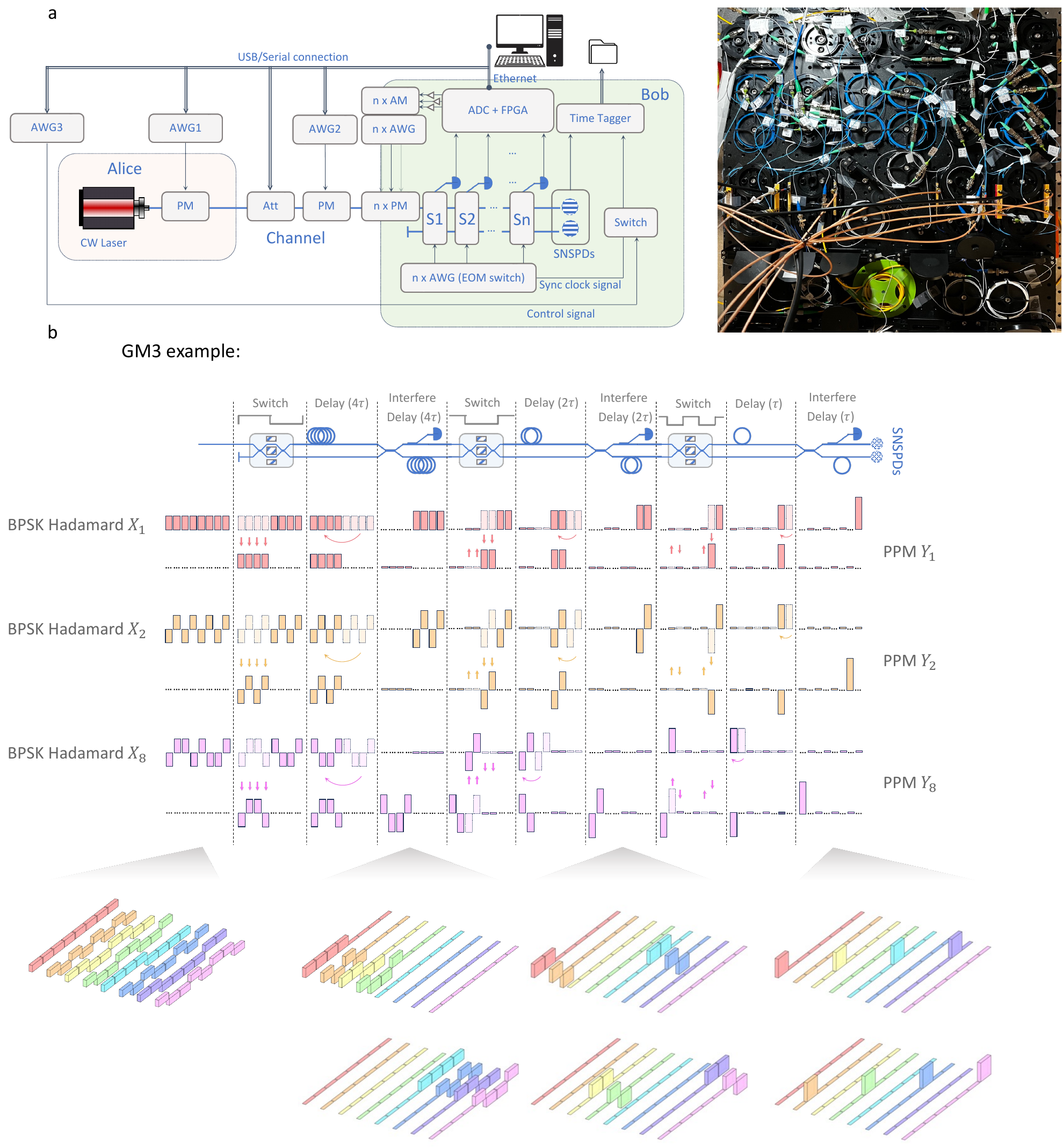}
    \caption{\textbf{a}, The experimental setups for testing Green Machines in this work. AWG: arbitrary wave generator. The synchronization clock signal is at 6.25 MHz for defining the package of each codeword when being registered by Time Tagger, which is also gated by a control signal from AWG3 to tell when the communication starts and ends. A photo of GM4 under construction is shown on the right. \textbf{b}, The logic flow for GM3 with details for Hadamard codeword 1, 2, and 8 through three stages of GM3. The delay at the lower arm after the beamsplitter is shown as moving the upper arm time bins forward. The intermediate patterns of all eight codewords after each stage are summarized at the bottom.}
    \label{fig:SuppS3}
\end{figure}

Following the Methods section at the end of the main text, here, we expand to the details of the experimental setup (\textbf{Supplementary Figure}~\ref{fig:SuppS3}\textbf{a}), the logic flows (\textbf{Supplementary Figure}~\ref{fig:SuppS3}\textbf{b}), and the time-multiplexing phase correction techniques (\textbf{Supplementary Figure}~\ref{fig:SuppS4}\textbf{a}) in this section. In these Supplementary Figures, we choose to give an example of GM3 since it is at the sweet spot between better visualization and enough complexity for generalization to more stages.

To test Green Machines under certain channel loss and phase drift, the experiments are repeated in multiple sets to reduce uncertainty caused by statistical fluctuation of limited photon detection events. Each set of experiments contains $N$ rounds of experiment for Alice to dial through all $N$ BPSK Hadamard codewords. Each round of the experiments is time-multiplexed into two phases. The first phase is phase correction. We need to correct and initial the phase of all the fiber interferometers in the Green Machine. The second phase is data acquisition. 

\subsection{A. Phase correction of the Green Machine}

\begin{figure}[htbp]
    \centering
    \vspace{-3pt}
    \includegraphics[width=5.5in]{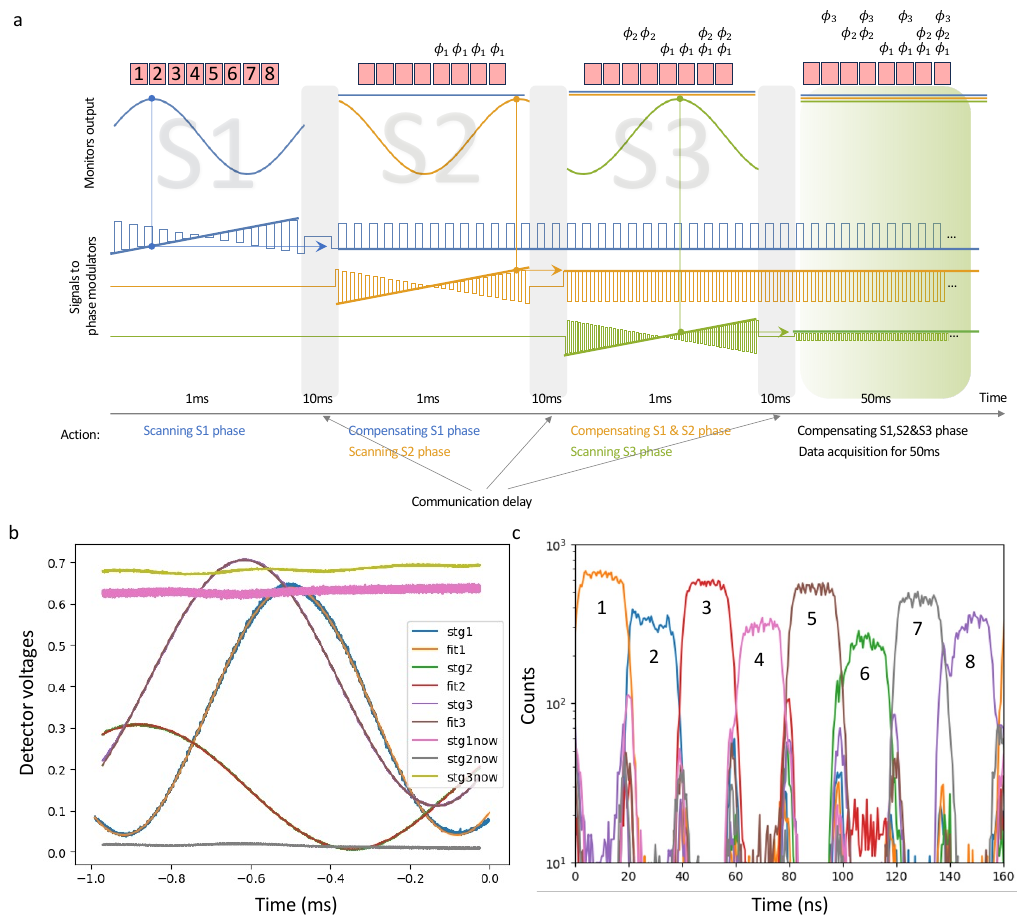}
    \caption{\textbf{a}, Timeline, monitor outputs, and signals for PMs during phase compensation and data acquisition of GM3. The modulated square waves' frequency is reduced for visualization. \textbf{b}, Compensation of phase drift by scanning phase and fitting the monitor voltages ($\sin$ waves). The near-straight lines tagged in ``stg now'' are the detectors' output voltage with a bright laser at around 100 ms after recording data with phase compensation to the maximum for S1, minimum for S2, and maximum for S3, which indicates that phase is corrected and stable during the data acquisition. \textbf{c}, An overlapped eight histograms for a piece of data showing measured 8-PPM codes transformed from BPSK Hadamard codes by GM3. Odd numbers are from the upper arm, and even numbers are from the lower arm.}
    \label{fig:SuppS4}
\end{figure}

Let's take GM3 as an example. During the phase correction of the GM3 (\textbf{Supplementary Figure}~\ref{fig:SuppS4}\textbf{a}), AWG1 and AWG2 are turned off. The channel attenuator is also set to zero attenuation so that the full-power CW laser is sent to Bob. Before the GM3 receiver, three concatenated phase modulators (PMs) are used to compensate for the phases of the three stages. The phase correction happens in serial, which means it first starts at S1. 

To know the current phase drift between two arms of S1, the FPGA generates a ramp signal that gets amplified to modulate the square wave amplitude of Bob's first AWG. The modulated signal is an amplitude-modulated square waveform at 6.25 MHz with a ramp-shape amplitude modulation with $V_\mathrm{pp}$ larger than $V_\pi$ of the PM and a period of 1 ms. This is to only add a scanning phase to the second half (5th, 6th, 7th, 8th) symbols of the codeword. Then, the photon detector monitoring one arm after the beamsplitter of the first stage will read an interference pattern as a $\sin$ function and send the pattern to the computer. After fitting the phase shift of the $\sin$ function, the computer will tell FPGA to generate a DC voltage to compensate for the S1 phase.  

The process for S2 and S3 is similar. The only difference is the frequency of the square waves. For S2, the square wave is at 12.5 MHz to apply an extra phase to the 3rd, 4th, 7th, and 8th symbols. For S3, it is at 25 MHz to add an extra phase to all even symbols (2nd, 4th, 6th, 8th). Note that the scan-and-fit phase compensation for the S2 phase must wait until the S1 phase is compensated and stable. For the S3 phase, it must wait until both S1 and S2 phases are compensated and stable. Otherwise, there will be ambiguity. 

The phase compensation points for all stages can be chosen as a combination of the maximum or minimum of the $\sin$ function, which only changes the code mapping. In \textbf{Supplementary Figure}~\ref{fig:SuppS4}\textbf{a}, the detailed flow shows the case where the phase compensation points are all maximum. 

We have tested the compensation performance by recording the monitor output with an additional $50$ ms delay after $50$ ms data acquisition and comparing it with the output during phase scanning. The comparison results of compensating to the maximum at S1, minimum at S2, and maximum at S3 are shown in \textbf{Supplementary Figure}~\ref{fig:SuppS4}\textbf{b}. A sampled piece of the histogram is shown in \textbf{Supplementary Figure}~\ref{fig:SuppS4}\textbf{c}. 

\subsection{B. Data acquisition}

Most of the details of the data acquisition, experimental logic, and data processing are explained in the main text and the Method. Here, we want to provide some additional information about how the data is recorded.

In the $j^\mathrm{th}$ round of one set of the experiment, the channel attenuation is set to the given value (through the serial port command from the computer) after completing phase correction of all stages in the Green Machine. To add channel phase drift, the AWG2 in \textbf{Supplementary Figure}~\ref{fig:SuppS3}\textbf{a} will send a corresponding signal to the channel PM. The channel parameter is loaded.

Then, Alice starts sending encoded Hadamard signals to her PM (by AWG1 in \textbf{Supplementary Figure}~\ref{fig:SuppS3}\textbf{a} through the serial port command from the computer) to encode the $j^\mathrm{th}$ BPSK Hadamard coherent states for 50 ms. After the command is applied, the computer will trigger AWG3 (through the serial port command) to send a high TTL control signal to the switch and let the 6.25 MHz clock signal pass to the Time Tagger. At the same time, the computer will ask Time Tagger to start recording the arrival time for both the SNSPD outputs and the rising edges of the clock signal. Therefore, by compensating for a constant time difference, the clock signal can define which $N$ serial of time bins forms one codeword.

After the Time Tagger returns the time-tagged events of all channels, the computer saves the file locally. The histogram of decoding each round of communication is then extracted from the corresponding file. From the histogram, the mean photon number received at the SNSPD and the transfer probability matrix for the given codeword are derived accordingly.

\begin{figure}[htbp]
    \centering
    \includegraphics[width=5in]{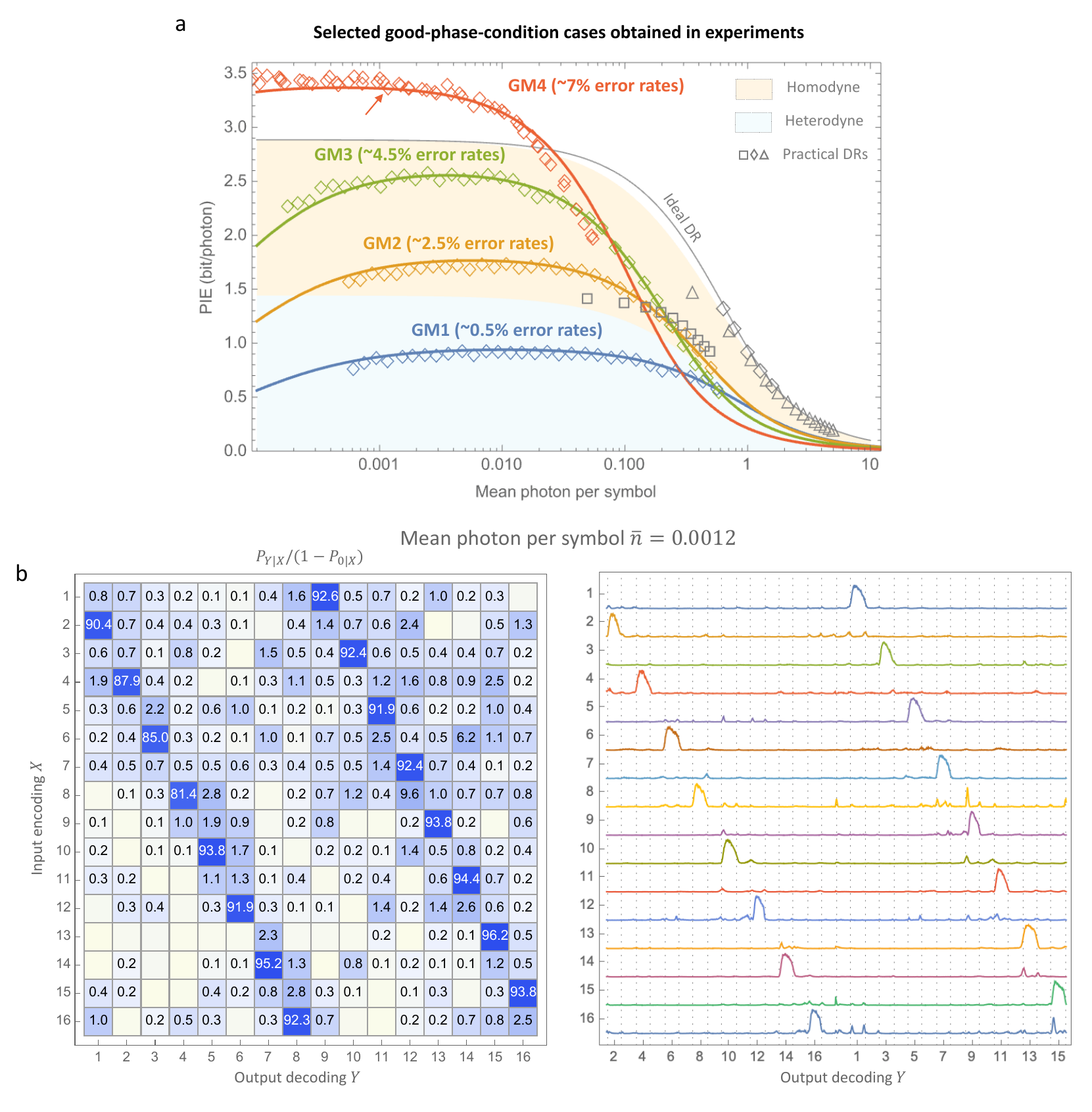}
    \caption{Stable cases selected based on the performance of phase correction (see \textbf{Supplementary Figure}~\ref{fig:SuppS4}\textbf{b}). \textbf{a}, Post-selected stable cases corresponding to tests at different mean photon numbers per symbol. \textbf{b}, The transition probability matrix and photon histogram of the red-arrow-pointed data set with $\bar{n}=0.0012$.}
    \label{fig:SuppS5}
\end{figure}

\subsection{C. Selected and non-selected data}

In Fig.2b of the main text, we show the non-selected data from multiple trials of all Green Machines, which even includes the data acquired when the Green Machine is occasionally in a bad phase condition. For example, the airflow and the daily activity in the lab sometimes cause unknown movement to the fibers and make the phase unstable. This can be seen from the second and eighth rows of histograms in Fig.2c of the main text, where more-than-usual crosstalk is recorded. However, since we have the monitor to check the performance of phase compensation and stability of the interferometers in the Green Machine, we are able to filter out the data acquired when the Green Machine is in a bad phase condition during the $50$ ms by confirming it with the monitor outputs (see \textbf{Supplementary Figure}~\ref{fig:SuppS4}\textbf{b}). Then, we can discard it and wait for the next round of testing the same codeword.

By doing so, we can extract the data that is not severely affected by the bad phase conditions while continuously running the Green Machine experiments. As shown in \textbf{Supplementary Figure}~\ref{fig:SuppS5}\textbf{a}, the crosstalks are reduced to 0.5\%, 2.5\%, 4.5\%, and 7.0\% for GM1, GM2, GM3, and GM4 accordingly. Compared to Fig.2b in the main text, this infers how well our Green Machines can work if the long-term phase stability is improved. We also include one example of GM4 at $\bar{n}=0.0012$ in \textbf{Supplementary Figure}~\ref{fig:SuppS5}\textbf{b} to show that low error rates can be obtained for all 16 codewords.

\textbf{The data processing and the generation of data-related supplementary figures in this work are done using Mathematica, Version 14.2, (2024)~\cite{Mathematica}.}

\bibliography{Ref}

\end{document}